\documentclass[prb,twocolumn,showpacs,superscriptaddress,preprintnumbers]{revtex4}

\usepackage{graphicx}
\usepackage{dcolumn}
\usepackage{amsmath}
\usepackage{color}

\begin{document}

\title{Metal-insulator transition and lattice instability of paramagnetic V$_2$O$_3$}

\author{I. Leonov}
\affiliation{Theoretical Physics III, Center for Electronic Correlations and Magnetism,
Institute of Physics, University of Augsburg, 86135 Augsburg, Germany}
\author{V. I. Anisimov}
\affiliation{Institute of Metal Physics, Sofia Kovalevskaya Street 18, 620990 Yekaterinburg
GSP-170, Russia}
\affiliation{Department of Theoretical Physics and Applied Mathematics, Ural Federal University, 620002 Yekaterinburg, Russia}
\author{D. Vollhardt}
\affiliation{Theoretical Physics III, Center for Electronic Correlations and Magnetism,
Institute of Physics, University of Augsburg, 86135 Augsburg, Germany}

\date{\today}

\begin{abstract}

We determine the electronic structure and phase stability of paramagnetic V$_2$O$_3$ at the
Mott-Hubbard metal-insulator transition (MIT) by employing a combination of an \emph{ab initio}
method for calculating band structures with dynamical mean-field theory. The structural transformation 
associated with the MIT occurs upon a slight expansion of the lattice volume by $\sim 1.5$ \%, in agreement 
with experiment. Our results show that the structural transition precedes the MIT, implying a complex
interplay between electronic and lattice degrees of freedom.
The MIT is found to be driven by a strong correlation-induced, orbital-selective renormalization of 
the V $t_{2g}$ bands. The effective electron mass of the $e_g^{\pi}$ orbitals diverges at the MIT.
Our results show that full charge self-consistency is crucial for a correct description of the physical 
properties of V$_2$O$_3$.

\end{abstract}

 \maketitle

\section{Introduction}

The question concerning the nature of a Mott metal-insulator phase transition (MIT) poses
one of the most fundamental problems in condensed-matter physics. It becomes an even greater
challenge in the case of materials with a strong interplay between electronic correlations
and lattice degrees of freedom. The most famous example is V$_2$O$_3$ which has generally
been considered to be the classical example of a material with a Mott-Hubbard MIT \cite{M68}.
Under ambient conditions, V$_2$O$_3$ is a paramagnetic metal with a corundum crystal structure
(space group $R\bar{3}c$) \cite{MR69,Antiferro}. Upon doping with Cr it undergoes a phase
transition to a paramagnetic Mott-Hubbard insulator without change of crystal symmetry \cite{MRR69}.
The transition from the paramagnetic insulator (PI) to the paramagnetic metal (PM) can also be
triggered by doping with Cr, which essentially amounts to the application of a negative pressure.
The MIT is intimately linked with an abrupt expansion of the lattice volume by $\frac{\Delta V}{V}\sim 1.3$ \%
and a reduction of the $c/a$ lattice parameter ratio by $\sim 1.4$ \% across the MIT, indicating
a strong coupling between electronic and lattice degrees of freedom \cite{MRR69}. In spite of
intensive research over several decades, an explanation of this mutual influence of
electronic structure and phase stability of paramagnetic V$_2$O$_3$ at the MIT is still missing.
Therefore investigations of V$_2$O$_3$ continue with a high level of activity \cite{DD14,GP12}.

A realistic description of V$_2$O$_3$ requires taking into account the interplay
of electronic correlations and lattice degrees of freedom on a microscopic level.
State-of-the-art methods for the calculation of the band structure, which allow for
\emph{ab initio} computations of the electronic and structural properties,
can account for neither the MIT nor the paramagnetic
insulating phase of V$_2$O$_3$ \cite{M94}.
This obstacle can be overcome by employing the LDA/GGA+DMFT approach \cite{LDA+DMFT},
a combination of band-structure methods in the local-density approximation (LDA) or
the generalized-gradient approximation (GGA) with dynamical mean-field theory (DMFT)
of strongly correlated electrons \cite{DMFT}.
Applications of LDA+DMFT have already provided a good quantitative description of the electronic
structure and spectral properties of V$_2$O$_3$ \cite{HK01,PA06,PT07,BP08,GP12,MD03,LB10}.
In particular, it was found that in all phases the V$^{3+}$ ions are in a $S=1$ spin configuration,
with a mixed ($a_{1g}$, $e_g^{\pi}$) and ($e_g^{\pi}$, $e_g^{\pi}$) orbital occupation.
These calculations also allow one to perform a direct comparison of the calculated spectra
with, e.g., photoemission, x-ray absorption, and optical conductivity measurements near
the metal-insulator transition \cite{HK01,PA06,PT07,BP08,GP12,MD03,LB10}.
The LDA+DMFT results capture all generic aspects of a Mott-Hubbard metal-insulator
transition, such as a coherent quasiparticle behavior, formation of the lower- and
upper-Hubbard bands, and strong renormalization of the effective electron mass.
In addition, these calculations reveal an orbital-selective behavior of the electron
coherence and a strong enhancement of the crystal-field splitting between the $a_{1g}$
and $e_g^{\pi}$ bands, caused by electron correlations at the MIT \cite{HK01,PT07}.
Despite this success the coupling between electronic correlations and lattice structure
at the Mott-Hubbard MIT in V$_2$O$_3$ is still poorly understood.
We will address this problem in our investigation and thereby shed new light on the long-standing
question regarding the origin of the structural changes in the vicinity of the MIT.

\section{Method}

In this paper we employ the GGA+DMFT computational approach \cite{LB08} to explore the electronic
and structural properties of paramagnetic V$_2$O$_3$ across the Mott-Hubbard MIT. In particular,
we will explore the structural phase stability of paramagnetic V$_2$O$_3$, i.e., the influence
of electronic correlations on the \emph{structural} transition.
We first compute the electronic structure of paramagnetic V$_2$O$_3$ within the nonmagnetic
GGA using the plane-wave pseudopotential approach \cite{PSEUDO}. To investigate the structural
stability, we use the atomic positions and the c/a lattice parameter ratio of V$_2$O$_3$ taken
from experiment \cite{MRR69}. To this end, we adopt the crystal structure data for paramagnetic
metallic V$_2$O$_3$ and insulating (V$_{0.962}$Cr$_{0.038}$)$_2$O$_3$, respectively,
and calculate the total energy as a function of volume.
In Fig.~\ref{fig_1} (inset) we show the results of the nonmagnetic GGA total energy
calculations, which agree with previous band structure results \cite{M94}.
In particular, we find equilibrium lattice constants $a = 4.923$ \AA\ for the PM phase and
4.955 \AA\ for the PI phase of V$_2$O$_3$. The calculated bulk moduli are $B \sim 252$ GPa
for both phases.
These results are in neither quantitative nor even qualitative agreement with experiment.
Namely, they give a metallic solution with no structural phase transition between the PM and PI phases.
Clearly, standard band-structure techniques cannot explain the properties of
paramagnetic V$_2$O$_3$ since they do not treat electronic correlations adequately.

\section{Results and discussion}


\subsection{Non-self-consistent calculations}

Therefore we now compute the electronic structure and phase stability of V$_2$O$_3$ using the GGA+DMFT
computational scheme \cite{LB08}. For the partially filled V $t_{2g}$ orbitals, which split
into $a_{1g}$ and $e_g^{\pi}$ bands due to a trigonal distortion, we construct a basis of
atomic centered symmetry-constrained Wannier functions \cite{AK05}. To solve the realistic
many-body problem within DMFT, we employ the continuous-time hybridization-expansion quantum
Monte-Carlo algorithm \cite{ctqmc,extra}. The calculations are performed at the temperature $T \sim$ 390 K,
which is below the critical endpoint of $T_c \sim 458$ K \cite{JM70}. We use values of the Coulomb
interaction $U = 5$ eV and Hund's exchange $J = 0.93$ eV, in accordance with the
previous theoretical and experimental estimations \cite{HK01,PA06,PT07,GP12}. The $U$ and $J$ values are
assumed to remain constant across the phase transition,
which is consistent with recent hard x-ray photoemission experiments \cite{FS11}.
Furthermore we employ the fully-localized
double-counting correction, calculated from the self-consistently determined local occupancies, to
account for the electronic interactions already described by the GGA.

\begin{figure}[tbp!]
\centerline{\includegraphics[width=0.4\textwidth,clip=true]{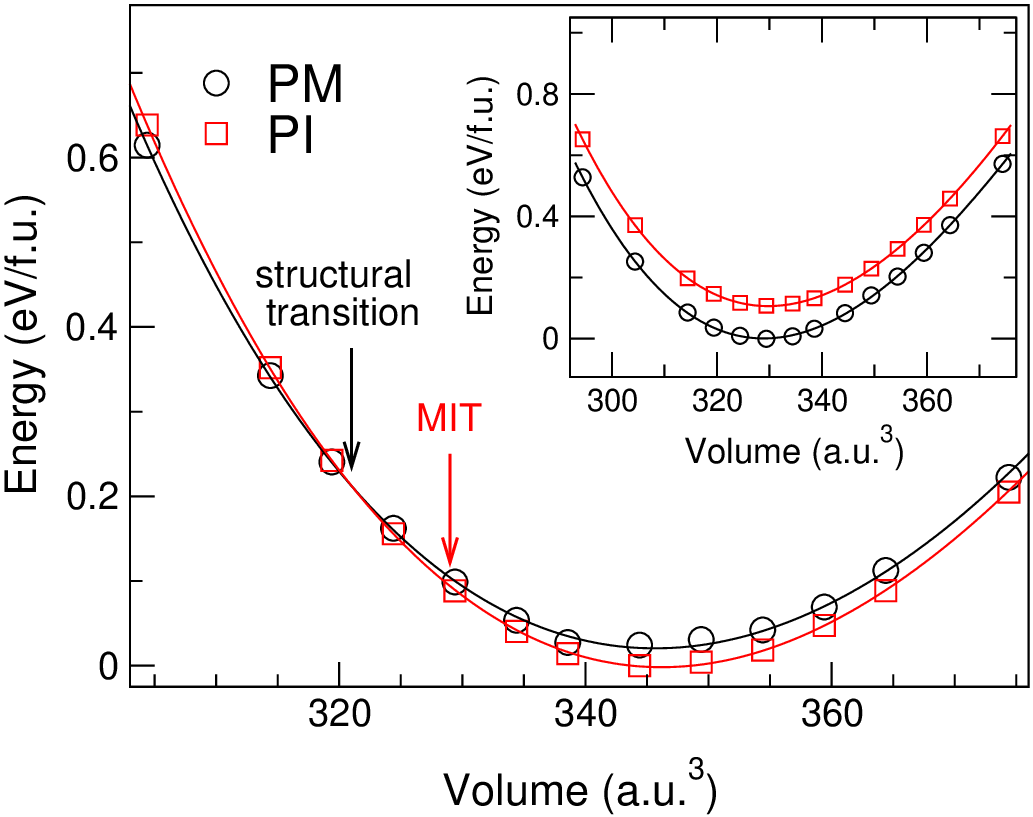}}
\caption{(Color online)
Variation of the total energy of paramagnetic V$_2$O$_3$ computed within GGA+DMFT for different volumes.
The inset shows the results of the nonmagnetic GGA total-energy calculations. }
\label{fig_1}
\end{figure}

In Fig.~\ref{fig_1} (main panel) we display the calculated variation of the total energy of paramagnetic
V$_2$O$_3$ as a function of lattice volume. Our results for the equilibrium lattice constant and bulk
modulus, which now include the effect of electronic correlations, agree well with experiment \cite{MRR69}.
The calculated equilibrium lattice constants are $a=$ 5.005 and 5.035 \AA\ for the PM and PI phases,
respectively. The corresponding bulk moduli are $B=$ 202 and 222 GPa.
In agreement with previous studies \cite{HK01,PT07}, our results show an orbital-dependence of
the electron coherence, with coherent quasiparticle behavior for the $a_{1g}$ orbital, while the
$e_g^{\pi}$ bands remain incoherent, with a large imaginary part of the self-energy at $T \sim$ 390 K.
We also evaluated the spectral function of paramagnetic V$_2$O$_3$ (not shown here) and determined the
MIT phase boundary.
The MIT is found to take place at a positive pressure of $p_{el} \sim$ 125 kbar, implying a $\frac{\Delta V}{V} \sim 5$ \%
reduction of the lattice volume. We find that both the
$a_{1g}$ and the $e_g^{\pi}$ quasiparticle weights remain finite at the MIT; that is, there is \emph{no}
divergence of the effective electron mass at the transition as it would occur in a Brinkman-Rice
picture of the MIT \cite{BR71}. Thus we conclude that the MIT is driven by
a strong enhancement of the $a_{1g}$--$e_g^{\pi}$ crystal-field splitting caused by electron
correlations \cite{HK01,PT07}.

The GGA+DMFT results are qualitatively different from those obtained with the nonmagnetic GGA and
provide clear evidence for a structural phase transition. However, this transition is found
to occur at a critical pressure of $p_c \sim$ 186 kbar, corresponding to
a large ($\sim 7$ \%) reduction of the lattice volume.
In addition, the results imply that the PM phase is energetically unfavorable at ambient pressure,
i.e., thermodynamically unstable, with a total energy difference with respect to the PI phase
of $\Delta E \equiv E_\mathrm{PM} - E_\mathrm{PI} \sim$ 20 meV/f.u.. These features are in contrast to
experiment.
The origin of this discrepancy can be ascribed to the lack of charge self-consistency
in the present calculations. Indeed, a strong enhancement of the $a_{1g}$--$e_g^{\pi}$
crystal-field splitting will cause a substantial redistribution of the charge density and
thereby influence the lattice structure due to electron-lattice coupling. All this makes
charge self-consistency \cite{charge-sc-LDA+DMFT} particularly important at the
metal-insulator transition.

\subsection{Self-consistent calculations}

\begin{figure}[tbp!]
\centerline{\includegraphics[width=0.4\textwidth,clip=true]{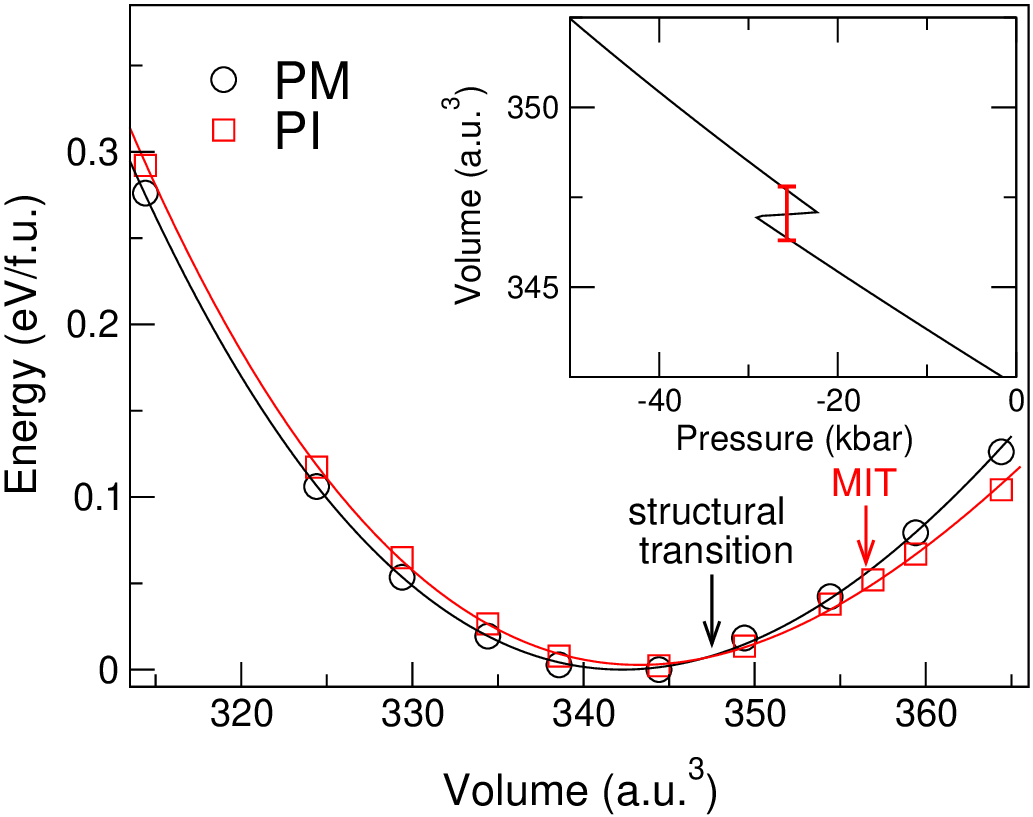}}
\caption{(Color online)
Variation of the total energy of paramagnetic V$_2$O$_3$ as a function of volume
obtained by employing the fully charge self-consistent GGA+DMFT method.
The inset shows the corresponding volume-pressure equation of state obtained as a derivative of the
spline interpolation of $E(V)$. The calculated critical pressure $p_c \sim -28$ kbar and
volume collapse by $\sim 0.5$ \% are marked by a red bar.
}
\label{fig_2}
\end{figure}

For this reason we implemented a fully charge self-consistent GGA+DMFT method \cite{charge-sc-LDA+DMFT}
within a plane-wave pseudopotential approach to compute the electronic structure and phase
stability of V$_2$O$_3$.
In Fig.~\ref{fig_2} we present our results for the total energy for different volumes.
The calculated pressure-volume equation of state is shown in Fig.~\ref{fig_2} (inset).
The PM phase is now found to be thermodynamically stable at ambient pressure,
with a total-energy difference between the PM and PI phases of $\Delta E \sim$ 3 meV/f.u.
The structural transition takes place upon a slight expansion of the lattice,
$\frac{\Delta V}{V}\sim$ 1.5 \%,
at a negative critical pressure of $p_c \sim -28$~kbar, in agreement with experiment. The phase transition
is accompanied by an abrupt increase of the lattice volume by $\sim$ 0.5 \% and a simultaneous change
of the $c/a$ ration by $\sim$ 1.5 \%. This result is seen to differ significantly from that obtained with the
non-charge-self-consistent GGA+DMFT scheme, according to which the PM phase is thermodynamically unstable at
ambient pressure.
The calculated equilibrium lattice constants and bulk moduli are now in remarkably good
agreement with experiment \cite{MRR69}. In particular, we obtain $a = 4.99$ and
5.021 \AA\ for the PM and PI phases, respectively, which is less than $1$ \% larger than
the experimental values. The calculated bulk moduli are $B = 219$ and 204 GPa,
respectively. We note that the bulk modulus in the PI phase is somewhat smaller than
that in the PM phase, which implies an \emph{enhancement} of the compressibility at
the phase transition.
Furthermore, the total-energy calculation results exhibit a weak anomaly near the MIT,
which is associated with a divergence of the compressibility, in accordance with
previous model calculations \cite{KM02}.
Overall, the electronic structure, the equilibrium lattice constant, and the structural phase
stability of paramagnetic V$_2$O$_3$ obtained with the fully charge self-consistent GGA+DMFT scheme
are in remarkably good agreement with the experimental data.
Our calculations clearly demonstrate the crucial importance of electronic correlations
and full charge self-consistency to explain the thermodynamic stability of the paramagnetic
metal phase of V$_2$O$_3$.

\subsection{Spectral function}

\begin{figure}[tbp!]
\centerline{\includegraphics[width=0.4\textwidth,clip=true]{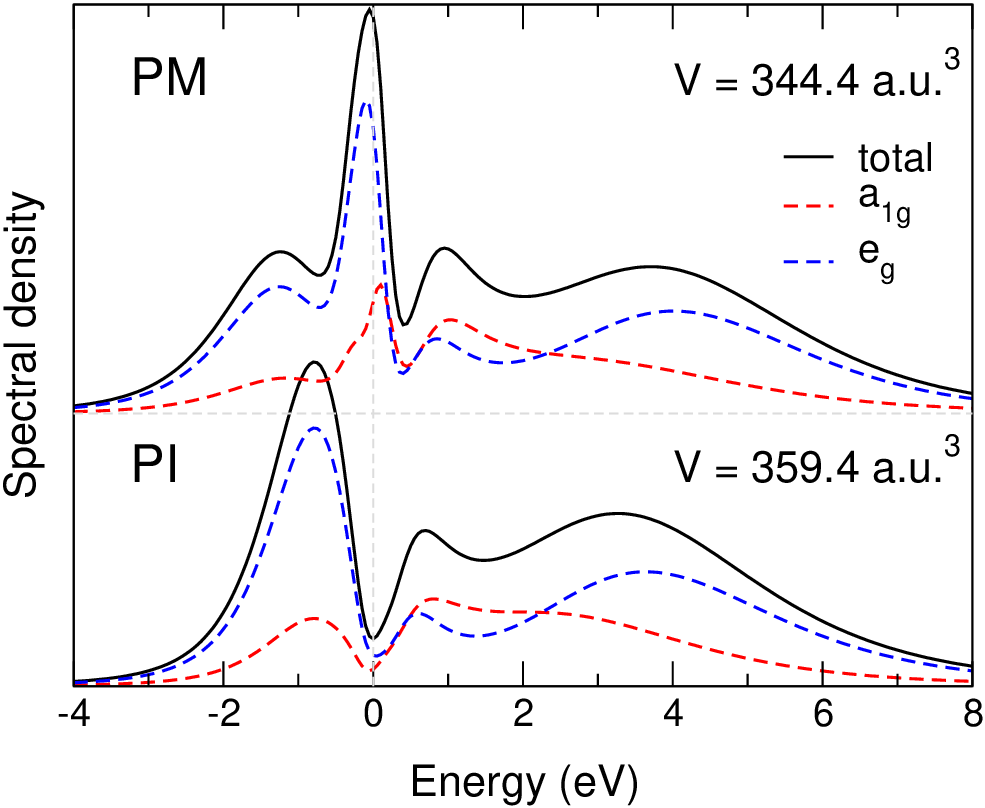}}
\caption{(Color online)
Spectral function of paramagnetic V$_2$O$_3$ across the MIT computed within GGA+DMFT.
}
\label{fig_3}
\end{figure}

In addition, we investigated the evolution of the spectral function of paramagnetic V$_2$O$_3$.
In Fig.~\ref{fig_3} we present our results obtained with the fully charge self-consistent
GGA+DMFT approach for paramagnetic V$_2$O$_3$ across the MIT.
Our calculations show that the V$^{3+}$ ions are in a $S=1$ spin configuration in all phases,
with a predominant occupation of the $e_g^{\pi}$ bands and substantial admixture of the $a_{1g}$
orbital. The admixture is almost independent of changes of the lattice volume (deviations $<5$ \%),
with a small decrease at the structural transition. Our results for the ($a_{1g}$, $e_g^{\pi}$) orbital
occupations at the phase transition are (0.44, 0.78) for the PM phase and (0.42, 0.79) for the PI phase
of V$_2$O$_3$. These findings are in good agreement with previous experimental estimations \cite{PT00}.
We note that the MIT takes place at a negative pressure
$p_{el} \sim -66$ kbar, i.e., upon an expansion of the lattice volume by $\sim$ 2 \%,
above the structural phase transition.
Thereby we conclude that the structural transition and the electronic transition are \emph{decoupled}.
Apparently the structural transformation occurs as a precursor to the Mott-Hubbard
MIT \cite{PO06}, implying an intricate interplay between electronic
and lattice degrees of freedom at the transition.

\subsection{Qusiparticle weight}

\begin{figure}[tbp!]
\centerline{\includegraphics[width=0.4\textwidth,clip=true]{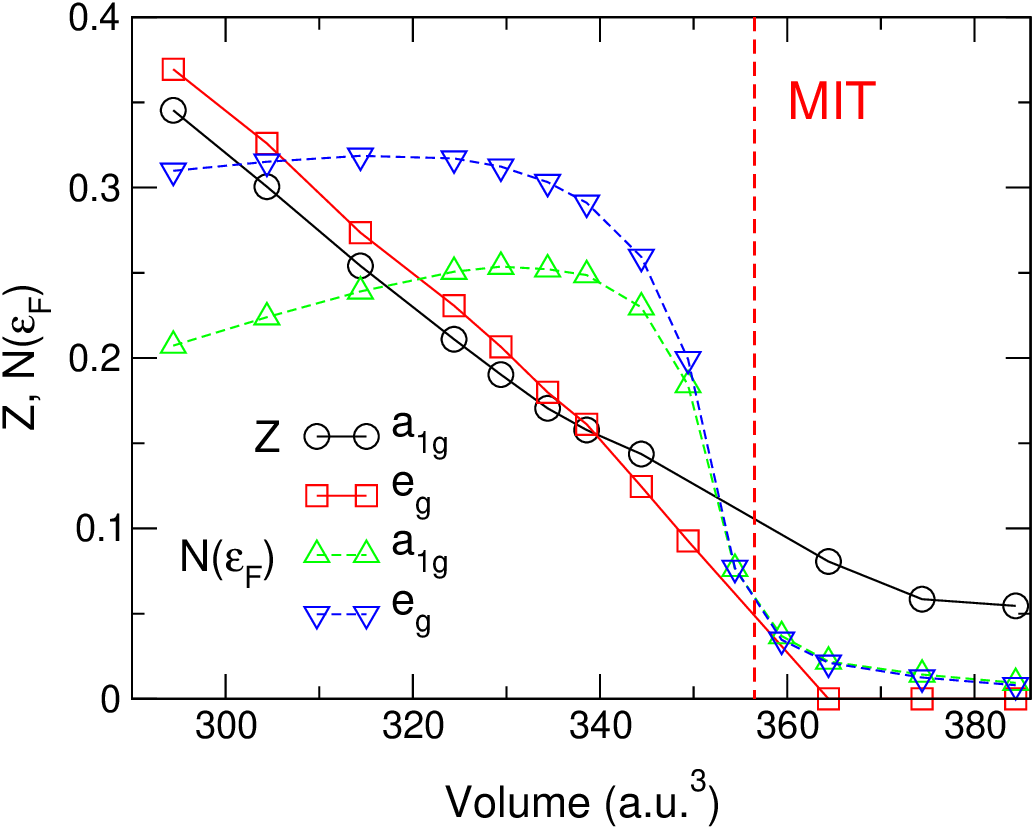}}
\caption{(Color online)
Quasiparticle weight $Z$ and spectral weight at the Fermi level
$N(\epsilon_F) = -\frac{\beta}{\pi} G(\tau=\beta/2)$ obtained for the $a_{1g}$ and
$e_g^{\pi}$ orbitals across the MIT in the PI phase
of V$_2$O$_3$. The MIT itself is indicated by a red dashed line.}
\label{fig_4}
\end{figure}

Finally, we calculate the quasiparticle weight employing a polynomial fit of the imaginary
part of the self-energy Im$\Sigma(i\omega_n)$ at the lowest Matsubara frequencies $\omega_n$.
It is evaluated as $Z=(1-\partial \textrm{Im} \Sigma(i\omega)/ \partial i\omega)^{-1}$ from
the slope of the polynomial fit at $\omega=0$. In Fig.~\ref{fig_4} we present our results for
the $a_{1g}$ and $e_g^{\pi}$ quasiparticle weights evaluated for the PI phase.
Upon lattice volume expansion, both the $a_{1g}$ and $e_g^{\pi}$ quasiparticle weights
monotonously decrease. Moreover, in the vicinity of the MIT the $a_{1g}$ quasiparticle
weight remains finite, while $Z = 0$ for the $e_g^{\pi}$ orbitals, i.e., Im$\Sigma(\omega)$
diverges for $\omega \rightarrow 0$. Therefore, we conclude that the electronic effective
mass of the $e_g^{\pi}$ bands diverges at the MIT, in agreement with thermodynamic
measurements of V$_2$O$_3$ \cite{MR71}. We note that this divergence coincides with
the drop of the spectral weight for the $a_{1g}$ and $e_g^{\pi}$ orbitals at the Fermi
level shown in Fig.~\ref{fig_4}.
The MIT is therefore driven by a strong orbital-selective renormalization of the V $t_{2g}$
bands, in accordance with a Brinkman-Rice picture of the MIT. These results correct previous
reasonings based on a correlation-induced enhancement of the crystal-field splitting of the
V $t_{2g}$ manifold, which results in a suppression of the hybridization between the $a_{1g}$ and
$e_g^{\pi}$ bands \cite{HK01,PT07}. On this basis, we conclude that an orbital-selective Mott
transition is the only viable scenario for the MIT in V$_2$O$_3$.

\section{Conclusion}

In conclusion, we employed the GGA+DMFT computational approach to determine the
electronic structure and phase stability of paramagnetic V$_2$O$_3$ across the Mott-Hubbard
metal-insulator phase transition. The calculated structural phase stability and spectral
properties are in good agreement with experiment. Full charge self-consistency is
found to be crucial to obtain the correct equilibrium lattice and electronic structure
of V$_2$O$_3$ at ambient pressure. Upon lattice expansion, a structural phase transition
is found to take place at $p_c \sim -28$ kbar, in agreement with experiment. The phase
transition is accompanied by an abrupt lattice volume expansion by $\sim$ 0.5 \% and a
simultaneous change of the $c/a$ lattice parameter ratio by about 1.5 \%. Our calculations
reveal an orbital-selective renormalization of the V $t_{2g}$ bands caused by strong electron
correlations. The electronic effective mass of the $e_g^{\pi}$ orbitals diverges at the MIT,
in accordance with a Brinkman-Rice picture of the MIT. However, the $a_{1g}$ orbital exhibits a
finite quasiparticle weight across the MIT.
We therefore conclude that an orbital-selective Mott transition is the only viable scenario
for the MIT in V$_2$O$_3$.
Most importantly we find that the structural transformation is decoupled from the electronic MIT.
Our findings highlight the subtle interplay between electronic correlations
and lattice stability across the Mott-Hubbard MIT.

\begin{acknowledgments}

We thank F. Lechermann and A. I. Poteryaev for valuable discussions. Support from the Deutsche
Forschungsgemeinschaft through TRR 80 (I.L.) and FOR 1346 (D.V.) is gratefully
acknowledged. V.I.A. acknowledges support from the Russian Scientific Foundation
(Project No. 14-22-00004), the Russian Foundation for Basic Research (Projects No.
13-02-00050), and the Ural Division of the Russian Academy of Science Presidium
(Project No. 15-8-2-4).

\end{acknowledgments}


\begin{thebibliography}{99}
\bibitem{extra}
We here employ the density-density approximation of the local Coulomb interaction between the V $t_{2g}$ electrons.
\end{thebibliography}
\end{document}